\documentclass[aps,prl,twocolumn,floatfix,showpacs,groupedaddress]{revtex4}
\usepackage{graphicx}
\usepackage{times}
\usepackage{amsbsy,amssymb,amsmath,bm}

\renewcommand{\vec}[1]{{\bm{#1}}}

\begin{document}
\title{ Circulating Current States in Bilayer Fermionic and Bosonic Systems}
\author{A. K. Kolezhuk}
\thanks{On leave from: Institute of
Magnetism, National Academy of Sciences and Ministry of Education,
03142 Kiev, Ukraine.}
\affiliation{Institut f\"ur Theoretische Physik C, RWTH Aachen, 52056 Aachen, Germany}
\affiliation{Institut f\"ur Theoretische Physik, Universit\"at Hannover, 30167  Hannover, Germany}

\begin{abstract}
It is shown that fermionic polar molecules or atoms in a bilayer optical lattice can
undergo the transition to a state with circulating currents, which
spontaneously breaks the time reversal symmetry. 
Estimates of relevant temperature scales are given and experimental signatures
of the circulating current phase are identified.
Related phenomena in bosonic
and spin systems with ring exchange are discussed.
\end{abstract}

\maketitle

\paragraph{Introduction.--}

The technique of ultracold  gases loaded into optical lattices
\cite{opt-lattice1,opt-lattice2} allows a direct experimental study of
paradigmatic models of strongly
correlated  systems. The possibility of
unprecedented control over the
model  parameters has opened wide perspectives for the study of quantum phase
transitions.  Detection of the Mott insulator to superfluid transition in bosonic atomic
gases \cite{opt-Hubbard,SF-Mott1,SF-Mott2},  of superfluidity 
\cite{Fermi-molecule-BEC,BEC-BCS} and Fermi liquid
 \cite{Fermi-surface} in cold Fermi gases, realization of Fermi systems with low
 dimensionality \cite{Fermi-lowd,Fermi-1d}
mark some of the recent achievements in this rapidly developing field \cite{Bloch-review}.  
 While the atomic interactions can be treated as 
contact ones for most purposes, polar
molecules \cite{Santos+00,Doyle+04,Buechler+07} could provide further opportunities of controlling
longer-range interactions.

In this Letter, I propose several models on a bilayer optical lattice which exhibit a
phase transition into an exotic circulating current state with
spontaneously broken time reversal symmetry. Those states are closely related to
the ``orbital antiferromagnetic states''  proposed first by Halperin
and Rice nearly 40 years ago \cite{HalperinRice68}, rediscovered two decades later
 \cite{AffleckMarston88,Schulz89,Nersesyan91} and recently found in numerical studies
in extended $t$-$J$ model on a ladder \cite{Schollwoeck+03} and  on a
two-dimensional bilayer \cite{Capponi+04}. Our goal
is to show how such states can be realized and detected in a relatively
simple optical lattice setup.

\paragraph{Model of fermions on a bilayer optical lattice.--}

\begin{figure}[b]
\begin{center}
\includegraphics[width=0.45\textwidth]{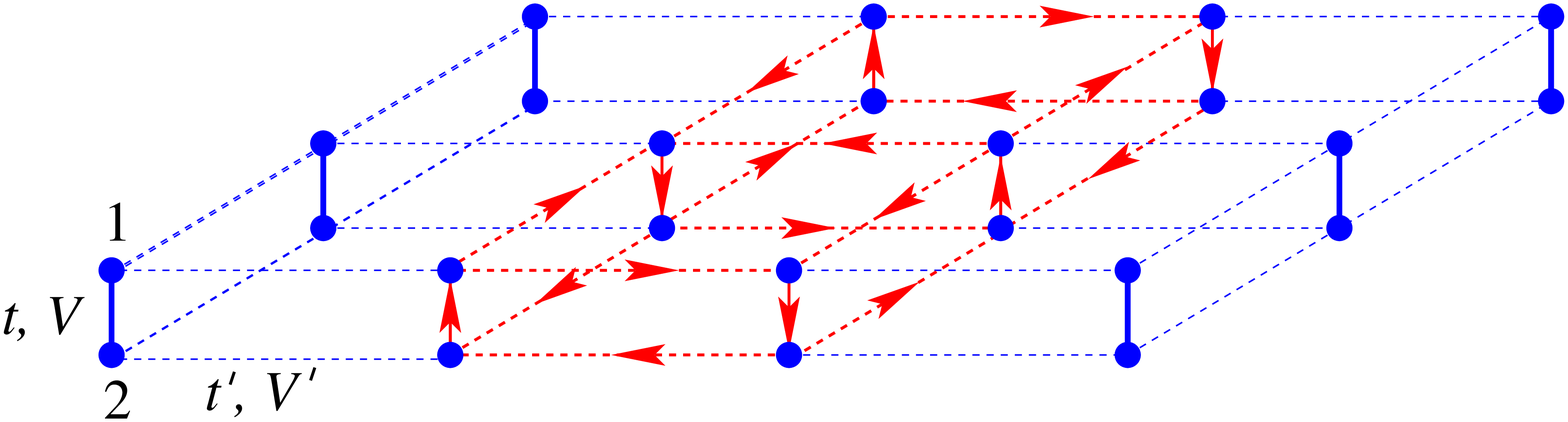}
\end{center}
\caption{\label{fig:bilayer} Bilayer lattice model described by the
  Hamiltonian (\ref{hamlat}). The arrows denote particle flow in the
  circulating current phase.
}
\end{figure}

Consider spin-polarized fermions in a bilayer  optical  lattice 
shown in Fig.\ \ref{fig:bilayer}. 
The system is described by the Hamiltonian
\begin{eqnarray} 
\label{hamlat} 
\mathcal{H}&=&V\sum_{\vec{r}}  n_{1,\vec{r}}n_{2,\vec{r}}
+\sum_{\sigma\sigma'} \sum_{\langle \vec{r}\vec{r}' \rangle}V'_{\sigma\sigma'}
n_{\sigma,\vec{r}}n_{\sigma',\vec{r}'}\\
&-&t\sum_{\vec{r}}(a^{\dag}_{1,\vec{r}}a^{\vphantom{\dag}}_{2,\vec{r}} +\mbox{h.c.})
-t'\sum_{\sigma}\sum_{\langle \vec{r}\vec{r}' \rangle}
(a^{\dag}_{\sigma,\vec{r}}
a^{\vphantom{\dag}}_{\sigma,\vec{r}'} +\mbox{h.c.}) \nonumber
\end{eqnarray}
where $\vec{r}$ labels the vertical dimers arranged in a two-dimensional (2d) square lattice,
$\sigma=1,2$ labels  two layers, and $\langle \vec{r}\vec{r}'\rangle$ denotes a sum over
nearest neighbors.  
Amplitudes $t$ and $t'$ describe hopping between the layers and within a
layer, respectively.
A strong ``on-dimer'' nearest-neighbor repulsion $V \gg
t,t'>0$ is assumed, and  there is an   interaction between the nearest-neighbor
dimers $V'_{\sigma\sigma'}$ which can be of either sign.

This seemingly exotic setup can be realized  by using
polar molecules \cite{Doyle+04,Buechler+07}, or atoms with a large dipolar magnetic moment such as $\rm
{}^{53}Cr$ \cite{Santos+00}, 
and adjusting the direction
 of the dipoles with respect to the bilayer
plane. Let $\theta$, $\varphi$  be the polar and azimuthal
angles of the dipolar moment (the coordinate axes  are 
along the basis vectors of the lattice,  $z$ axis is perpendicular to the
bilayer plane). Setting $\varphi=\pm\frac{\pi}{4},\pm\frac{3\pi}{4}$ ensures
the dipole-dipole
interaction is the same along the $x$ and $y$ directions. The
nearest neighbor interaction parameters in (\ref{hamlat}) take the
following values: 
$V=(d_{0}^{2}/\ell_{\perp}^{3})(1-3\cos^{2}\theta)$,
and 
$ V'_{12}=V'_{21}=(d_{0}^{2}/R^{3})\{
1-3R^{-2}(\ell_{\parallel}\cos\theta+\ell_{\perp}\sin\theta\cos\varphi)^{2}
\}$, $V'_{11}=V'_{22}=V'_{12}(\ell_{\parallel}=0)$, where
$d_{0}$ is the dipole moment of the particle, $\ell_{\perp}$ and
$\ell_{\parallel}$ are the lattice spacings in the directions
perpendicular and parallel to the layers, respectively, and
$R^{2}=\ell_{\parallel}^{2}+\ell_{\perp}^{2}$. 
The strength and the sign of interactions $V$, $\widetilde{V}'$ can be controlled by
tuning the angles $\theta$, $\varphi$ and the lattice constants $\ell_{\perp}$,
$\ell_{\parallel}$. 
Below we will see that the physics of the problem depends on the
difference  
\begin{equation} 
\label{V'} 
\widetilde{V}'= V'_{11}- V'_{12},
\end{equation}
with the most interesting regime
corresponding to $\widetilde{V}'<0$.

Consider the model at half-filling.  Since $V\gg t,t'$, we may restrict
ourselves to the reduced Hilbert space containing only states
with one fermion per dimer. Two states of each dimer can be identified 
with pseudospin-$\frac{1}{2}$ states $|\!\!\uparrow\rangle$ and
$|\!\!\downarrow\rangle$.  Second-order perturbation theory in $t'$ yields the
effective Hamiltonian
\begin{eqnarray} 
\label{hamf} 
 \mathcal{H}_{S} &=&\sum_{\langle \vec{r}
  \vec{r}'\rangle} \big\{ J(S_{\vec{r}}^{x}S_{\vec{r}'}^{x}+S_{\vec{r}}^{y}S_{\vec{r}'}^{y})
+ J_{z} S_{\vec{r}}^{z}S_{\vec{r}'}^{z}\big\} -H\sum_{\vec{r}} S_{\vec{r}}^{x},\nonumber\\
 J&=&{4(t')^{2}}/{V},\quad J_{z}\equiv J\Delta=J+\widetilde{V}',\quad H=2t,
\end{eqnarray}
describing a 2d anisotropic Heisenberg antiferromagnet in a magnetic field
perpendicular to the anisotropy axis.  The twofold degenerate ground state  has the
N\'eel antiferromagnetic (AF) order transverse to the field, with spins canted
towards the field direction. The AF order is along the $y$ axis for $\Delta<1$
(i.e., $\widetilde{V}'<0$), and along the $z$ axis for $\Delta>1$
($\widetilde{V}'>0$).  The  angle $\alpha$ between the spins
and the field  is classically given by
$\cos\alpha=H/(2ZJS)$, where $S$ is the spin value and $Z=4$ is the lattice
coordination number.  
This classical ground state is
exact at the special point
$H=2SJ\sqrt{2(1+\Delta)}$
\cite{Kurmann+82}. 
The transversal AF order vanishes above a certain critical field
$H_{c}$; classically $H_{c}=2ZJS$, and the
same result follows from the spin-wave
analysis of (\ref{hamf})  (one starts with the fully
polarized spin state  at large $H$ and looks when the magnon
gap vanishes). This expression becomes exact at the isotropic point $\Delta=1$
and is a good approximation for $\Delta$ close to $1$.

\begin{figure}[b]
\begin{center}
\includegraphics[width=0.35\textwidth]{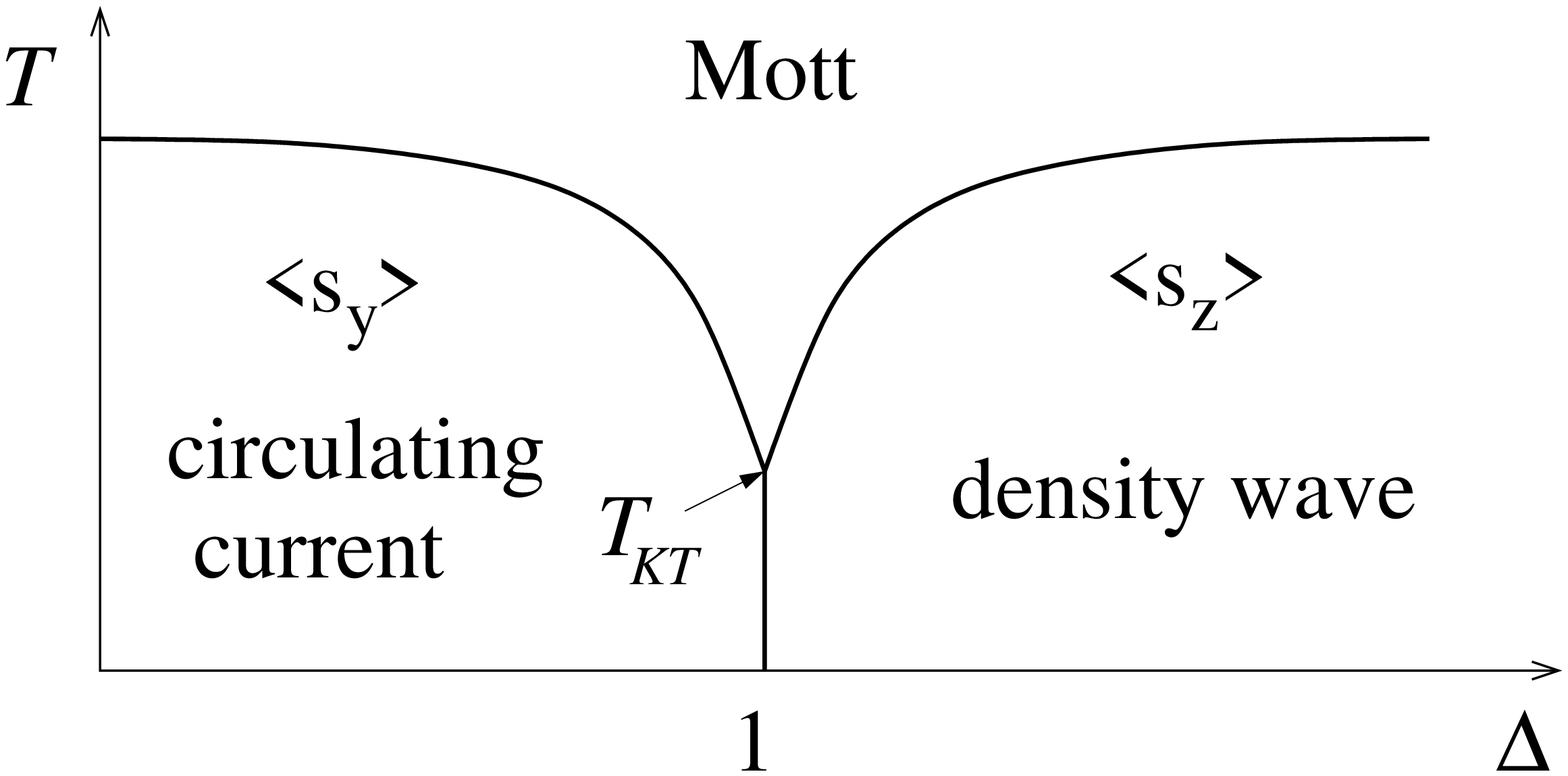}
\end{center}
\caption{\label{fig:fdiag} Schematic phase diagram of the model (\ref{hamlat}),
  (\ref{hamf}) at fixed  $H=2t$. The line $\Delta=1$ corresponds to
  the Kosterlitz-Thouless phase, with the transition temperature 
  $T_{KT}\propto J/\ln(J/H)$ at small $H$. 
}
\end{figure}

The long-range AF order along the $y$ direction translates in the
original fermionic language into the
staggered arrangement of \emph{currents} flowing from one layer to the other:
\begin{equation} 
\label{current} 
N_{y}=(-)^{r}\langle S_{r}^{y} \rangle \mapsto (-)^{r}\langle
-\frac{i}{2}(a^{\dag}_{1,r}a^{\vphantom{\dag}}_{2,r} -
a^{\dag}_{2,r}a^{\vphantom{\dag}}_{1,r})\rangle. 
\end{equation} 
In terms of the original
model (\ref{hamlat}), the condition $H<H_{c}$ for the existence of
such a staggered current  order becomes
\begin{equation} 
\label{hc} 
t<8(t')^{2}/V.
\end{equation}
The continuity equation for the current and 
the lattice symmetry dictate the current pattern
shown in Fig.\ \ref{fig:bilayer}. This circulating current (CC) state
has a spontaneously
broken time reversal symmetry, and is realized only for
attractive inter-dimer interaction $\widetilde{V}'<0$ (i.e.,
the easy-plane anisotropy $\Delta<1$) \cite{DuanDemlerLukin03}.  
If $\Delta=1$, the direction of the AF order in the $xy$
plane is arbitrary, so there is no long-range order at any finite
temperature. For $\Delta>1$ (i.e., $\widetilde{V}'>0$) the AF
order along the $z$ axis corresponds to the density wave (DW) phase  
with in-layer occupation numbers  having a finite staggered component.

The phase diagram in the temperature-anisotropy plane is sketched in Fig.\
\ref{fig:fdiag}.  
At the critical temperature $T=T_{c}$  
the discrete $\rm Z_{2}$ symmetry gets spontaneously broken, so the
corresponding thermal phase transition belongs to the 2d Ising
universality class (except the two lines $\Delta=1$ and $H=0$ 
where the symmetry is enlarged to $\rm U(1)$ and 
the transition becomes the Kosterlitz-Thouless one).  
Away from the phase boundaries the critical temperature
$T_{c}\sim J$, but at the isotropic point $\Delta=0$, $H=0$ it vanishes due to
divergent thermal fluctuations:
for $1-\Delta \ll 1$ and  $H\ll J$,
it can be estimated as
\begin{equation} 
\label{Tc}
T_{c}\sim J/\ln[\min(|1-\Delta|^{-1}, {J^{2}}/{H^{2}}) ].
 \end{equation}
The quantum phase
transition at $T=0$, $H=H_{c}$ is 
of the 3d Ising type (except at the $\rm U(1)$-symmetric point $\Delta=1$
where the universality class is that of the 2d dilute Bose gas
\cite{Subir-book}), so in its vicinity the CC order parameter
$N_{y}\propto (H_{c}-H)^{\beta}$ with $\beta\simeq 0.313$
\cite{3D-Ising}, and $T_{c}\propto J
N_{y}^{2}\propto J(H_{c}-H)^{2\beta}$. At $T>T_{c}$ or $H>H_{c}$ the only 
order parameter is $\langle S^{x}\rangle$, corresponding to the
Mott phase with one particle per dimer.

\paragraph{Bilayer lattice design and hierarchy of scales.--}

The bilayer can be realized, e.g., by employing three pairs of mutually
perpendicular counter-propagating laser beams with the same polarization and
adding another pair of beams with an orthogonal polarization and additional
phase shift $\delta$, so that the resulting field intensity has the form $\big[
E_{\perp}(\cos kx +\cos ky)+E_{z}\cos kz
\big]^{2}+\widetilde{E}_{z}^{2}\cos^{2}(kz+\delta)$. Taking
$\delta=\frac{\pi}{4}(1+\zeta)$ and
$\widetilde{E}_{z}^{2}>E_{z}(2E_{\perp}+\zeta E_{z})$, with $\zeta=\pm1$ for
blue and red detuning, respectively, one obtains a three-dimensional stack of
bilayers, separated by large potential barriers $U_{3d}$.  
Eq.\ (\ref{hc}) implies $V\gg t'\gg t,|\widetilde{V}'|$, which can be
achieved by making
the $z$-direction  potential barrier $U_{\perp}$ inside the
bilayer  sufficiently larger than
the in-plane barrier $U_{\parallel}$, so that the condition $t\ll t'$
will be met; e.g., $\widetilde{E}_{z}/E_{\perp}\approx 20$, $E_{z}/E_{\perp}\approx
15$ yields the barrier ratio $U_{3d} : U_{\perp}: U_{\parallel}$ of
approximately $16:8:1$, and the lattice constants $\ell_{\perp}\approx
0.45\lambda$, $\ell_{\parallel}=\lambda$, where $\lambda=2\pi/k$ is the laser wave length.
  The
parameter $\widetilde{V}'$ has a zero as a function of the angle $\theta$, so it
can be made as small as needed.
Taking 
$\lambda=400$~nm, 
one obtains an
estimate of  $T_{c}=(0.1\div 0.3)$~$\rm \mu K$ for cyanide molecules  $\rm
ClCN$ and $\rm HCN$ with the dipolar moment $d_{0}\approx
3$~Debye, while the
Fermi temperature for the same parameters is $T_{f}\approx (0.6\div 1.3)$~$\rm \mu K$.
This estimate
 corresponds to the maximum value of $T_{c}\sim J$ reached when
$\widetilde{V}'\sim -J$ and $t\lesssim J$. The hopping $t'$ was  estimated assuming
 the in-plane
 potential barrier $U_{\parallel}$ is roughly equal to the recoil energy $E_{r}=(\hbar
 k)^{2}/2m$, where $m$ is the particle mass.

\begin{figure}[tb]
\begin{center}
\includegraphics[width=0.23\textwidth]{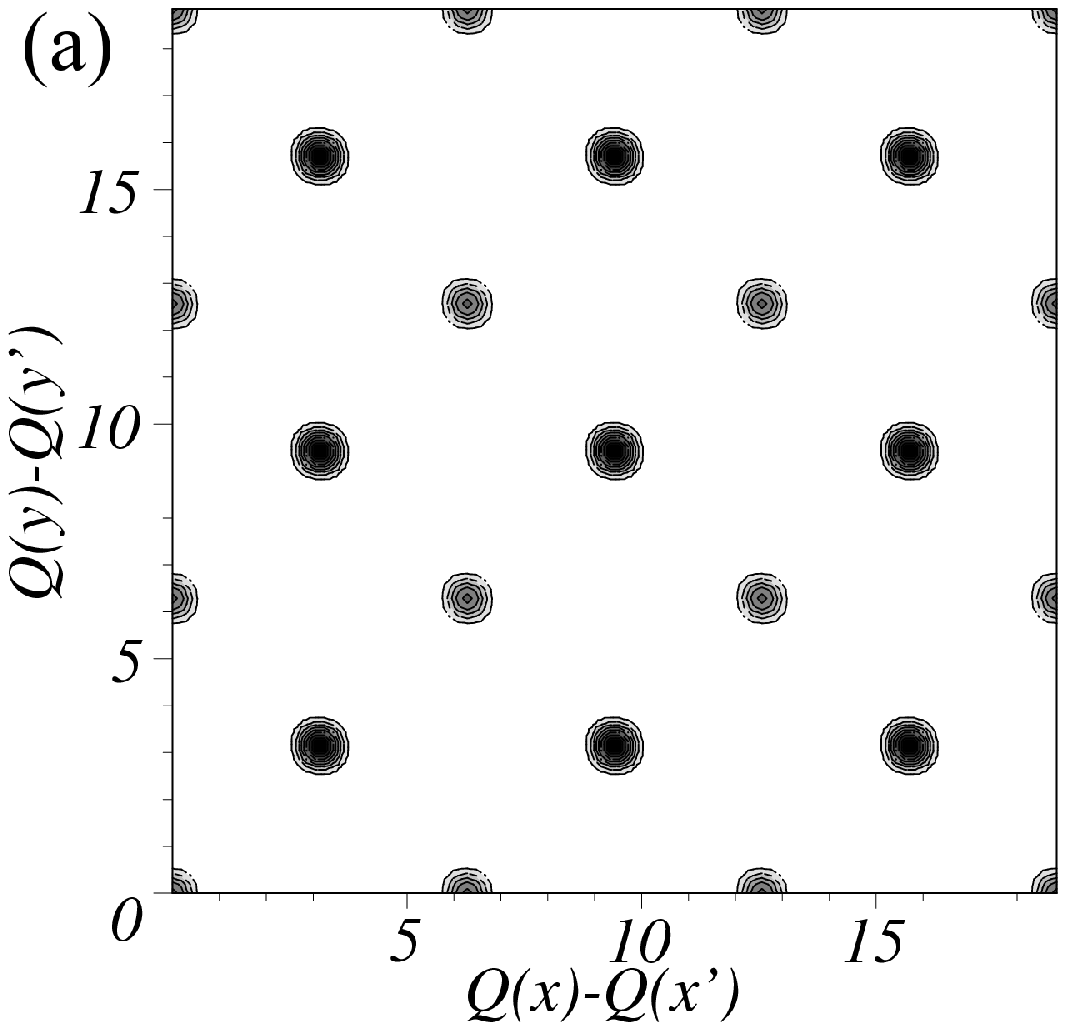}%
\includegraphics[width=0.23\textwidth]{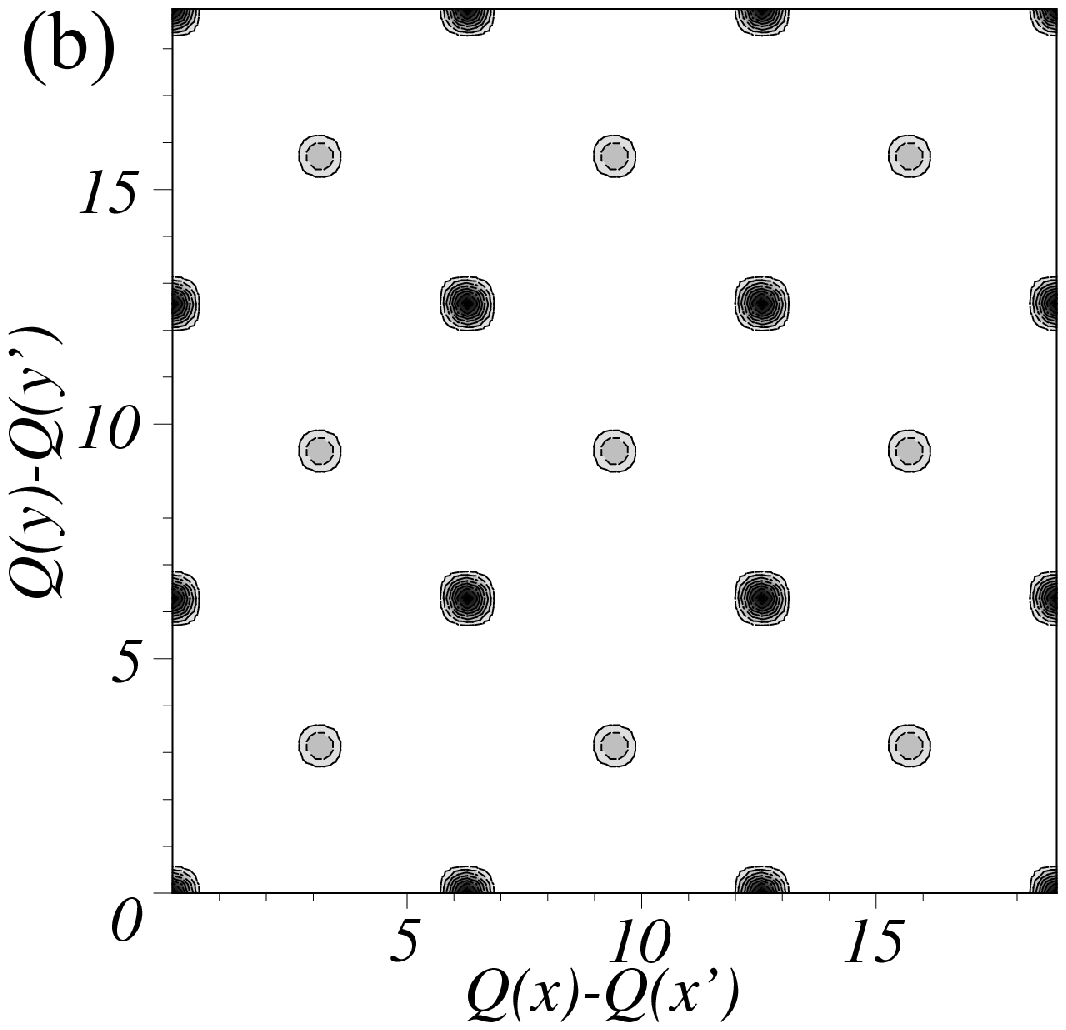}
\end{center}
\caption{\label{fig:tof} Noise
  correlation function
  $\mathcal{G}(\vec{r},\vec{r}')$ 
from time-of-flight images in the circulating current (CC) phase, shown as the
  function of the relative distance $\vec{Q}(\vec{r})-\vec{Q}(\vec{r}')$, with
  $\vec{Q}(\vec{r})=m\vec{r}/(\hbar t)$ expressed in
  $1/\ell_{\parallel}$ units: (a) $\vec{Q}(\vec{r})=(0,0)$; (b)
  $\vec{Q}(\vec{r})=\frac{\pi}{2}(1,1)$. 
  Changing the initial point $\vec{Q}(\vec{r})$ leads to the change of
  relative weight of the two systems of dips, which is the fingerprint of the CC phase.
}
\end{figure}

\paragraph{Experimental signatures. --}

Signatures of the ordered phases can be observed
 \cite{AltmanDemlerLukin04,Folling+05} in time-of-flight experiments
 by measuring the density noise correlator 
$\mathcal{G}(\vec{r},\vec{r'})=\langle n(\vec{r})n(\vec{r'})\rangle -\langle
n(\vec{r})\rangle \langle n(\vec{r'})\rangle$. If the imaging axis is
perpendicular to the bilayer, 
$n(\vec{r}) =\sum_{\sigma}\langle
a^{\dag}_{\sigma,\vec{r}}a^{\vphantom{\dag}}_{\sigma,\vec{r}}\rangle$ 
is the local net density of two layers. 
For large flight times $t$ it is proportional to the momentum 
distribution  $n_{\vec{Q(\vec{r})}} $, where
 $\vec{Q}(\vec{r})=m\vec{r}/\hbar t$. 
In the Mott  phase 
the response shows  fermionic ``Bragg dips'' at
 reciprocal lattice vectors $\vec{g}=({2\pi h}/{\ell_{\parallel}},
{2\pi k}/{\ell_{\parallel}})$,
\[ 
\mathcal{G}_{\rm M}(\vec{r},\vec{r'}) \propto 
f_{0}(\vec{r},\vec{r'}) = -2\langle S^{x}\rangle^{2}\sum_{\vec{g}}
 \delta\big( \vec{Q}(\vec{r}) -\vec{Q}(\vec{r'}) -\vec{g} \big) 
\]
 In the  CC and DW phases the noise
 correlator contains an additional system of dips
 shifted by  $\vec{Q}_{B}=({\pi}/{\ell_{\parallel}},
{\pi }/{\ell_{\parallel}})$:
\begin{eqnarray}
&& \mathcal{G}_{{\rm CC,DW}}(\vec{r},\vec{r'})\propto f_{0}(\vec{r},\vec{r'})
-2\Big\{ \langle S^{z}\rangle^{2} 
+\langle
S^{y}\rangle^{2}\nonumber\\ 
&&\quad \times \big[ 1   +\big(\cos\big(Q_{x}(\vec{r})\ell_{\parallel}\big) 
+\cos\big(Q_{y}(\vec{r})\ell_{\parallel}\big)\big)^{2}\big]\Big\} \nonumber\\
&&\qquad \times \sum_{\vec{g}}
 \delta\big( \vec{Q}(\vec{r}) -\vec{Q}(\vec{r'}) -\vec{Q}_{B}-\vec{g} \big) 
\end{eqnarray}
 In the DW phase $\langle S^{z}\rangle \not=0$, $\langle S^{y}\rangle=0$, and so
 the density correlator depends only on $\vec{r}-\vec{r'}$. In the CC phase
 $\langle S^{z}\rangle =0$, $\langle S^{y}\rangle\not=0$, and the relative
 strength of the two systems of dips varies periodically when one changes the
 initial point $\vec{r}$, see Fig.\ \ref{fig:tof}.  This
 $\vec{Q}$-dependent contribution stems from the intra-layer currents $\langle
 a^{\dag}_{\sigma,\vec{r}}a^{\vphantom{\dag}}_{\sigma,\vec{r'}}\rangle=
 (-)^{\sigma}(-)^{\vec{r}} \delta_{\langle \vec{r} \vec{r'}\rangle} i\langle
 S^{y}\rangle/4$, where  $\frac{1}{4}$ comes from the fact that the
 inter-layer current splits into four equivalent intra-layer ones (see Fig.\
 \ref{fig:bilayer}), $\delta_{\langle \vec{r} \vec{r'}\rangle}$ means 
$\vec{r}$ and $\vec{r'}$ must be nearest neighbors,
 and $(-)^{\vec{r}}\equiv e^{i\vec{Q}_{B}\cdot \vec{r}}$ denotes an
 oscillating factor.  If the 
 correlator is averaged over the particle positions, the CC and DW phases become
 indistinguishable. 
A  direct
 way to observe the CC phase could be to use the laser-induced
 fluorescence spectroscopy to detect the Doppler line splitting proportional to the
 current.

\paragraph{Bosonic models.--}

Consider the bosonic version of the model (\ref{hamlat}), with the
additional on-site repulsion $U$. The effective Hamiltonian has the
form (\ref{hamf}) with $J=-4(t')^{2}/V$ and
$J_{z}=\widetilde{V}'+4(t')^{2}(1/V-1/U)$.  Due to ferromagnetic (FM)
transverse exchange, 
instead of spontaneous current one obtains the usual
Mott phase.  CC states can be induced by artificial gauge fields
\cite{art-gauge}: The vector potential
$\vec{A}(x)=\frac{\pi}{\ell_{\parallel}}(x+1/2)$ makes hopping along
the $x$ axis imaginary, $t'\mapsto it'$. The unitary transformation
$S^{x,y}_{\vec{r}}\mapsto (-)^{\vec{r}}S^{x,y}_{\vec{r}}$ maps the
system onto a set of FM chains along the $x$ axis,
AF-coupled in the $y$ direction and subject to a staggered field
$H=2t$ along the $x$ axis in the easy $(xy)$ plane. In the ground
state net chain moments are arranged in a staggered way
along the $y$ axis, so a current pattern similar to that of Fig.\
\ref{fig:bilayer} emerges, now staggered along only one of the two
in-plane directions.

A different type of CC states, with orbital currents localized at lattice sites,
can be achieved with $p$-band bosons \cite{Wu+06}.

Yet another way to create a CC state in a bosonic bilayer
is to introduce the ring exchange on vertical plaquettes:
\begin{equation} 
\label{b-ring} 
\mathcal{H}_{\rm ring}=\frac{1}{2}K\sum_{\langle \vec{r}\vec{r'}\rangle}
(b^{\dag}_{1,\vec{r}}b^{\dag}_{2,\vec{r'}} b^{\vphantom{\dag}}_{2,\vec{r}}  
b^{\vphantom{\dag}}_{1,\vec{r'}} +\mbox{h.c.}).
\end{equation}
In  pseudospin language, the ring interaction 
modifies the transverse exchange, $J\mapsto J+K$, so for $K>0$
one can achieve the conditions $J>0$, $J>|J_{z}|$ necessary for the
CC phase to exist. However, engineering a sizeable ring exchange in
bosonic systems is difficult  (see \cite{Buechler+05} for recent proposals).

\paragraph{Spin-$\frac{1}{2}$ bilayer with four-spin ring exchange.-- }

Consider the Hubbard model for spinful fermions on a bilayer shown in Fig.\
\ref{fig:bilayer},
with the on-site repulsion $U$ and inter-
and intra-layer hoppings $t$ and $t'$, respectively. At half
filling (i.e., two fermions per dimer),
one can effectively describe the system in terms of spin degrees of freedom
represented by the operators
$\vec{S}=\frac{1}{2}a^{\dag}_{\alpha}\vec{\sigma}_{\alpha\beta}a^{\vphantom{\dag}}_{\beta}$. 
The leading term in $t/U$  yields the
AF Heisenberg model with the nearest-neighbor exchange constants
$J_{\perp}=4t^{2}/U$ (inter-layer) and $J_{\parallel}=4(t')^{2}/U$ (intra-layer), 
while the next term, with the interaction strength 
$J_{4}\simeq 10t^{2}(t')^{2}/U^{3}$,  corresponds
to the ring exchange \cite{Takahashi77,MacDonald+90}:
\begin{eqnarray} 
\label{s-ring}
\mathcal{H}_{4}&=&2J_{4}\sum_{\square}\big\{
(\vec{S}_{1}\cdot \vec{S}_{2}) (\vec{S}_{1'}\cdot \vec{S}_{2'})\nonumber\\
&+&(\vec{S}_{1}\cdot \vec{S}_{1'}) (\vec{S}_{2}\cdot \vec{S}_{2'})
-(\vec{S}_{1}\cdot \vec{S}_{2'}) (\vec{S}_{2}\cdot \vec{S}_{1'})
\big\},
\end{eqnarray}
where the sum is over vertical plaquettes only (the interaction for intra-layer
plaquettes is of the order of $(t')^{4}/U^{3}$ and is neglected), and
the sites $(1,2,2',1')$  form a plaquette (traversed counter-clockwise). 
In the same order of the perturbation theory, the nearest-neighbor exchange constants get
corrections, 
\[
J_{\perp}\mapsto J_{R}=J_{\perp}+J_{4}, \quad
J_{\parallel}\mapsto J_{L}=J_{\parallel}+J_{4}/2, 
\]
and the interaction $J_{D}=\frac{1}{2}J_{4}$ along the diagonals of vertical
plaquettes  is generated. Generalization for any 2d bipartite lattice built of
vertically arranged dimers is trivial.
Since $J_{\perp}\gg J_{\parallel}, J_{4}$, we
can treat the system as a set of weakly coupled spin dimers. 
The dynamics can be described with the help of the effective field
theory \cite{K96} which is a continuum
version of the bond boson approach \cite{SachdevBhatt90} and is based on
dimer coherent states 
 $|\vec{u},\vec{v}\rangle= (1-u^{2}-v^{2})|s\rangle
+\sum_j (u_{j}+iv_{j})|t_j\rangle$.
Here $|s\rangle$ and $|t_j\rangle$, $j=(x,y,z)$ are the singlet and 
triplet states, and $\vec{u}$, $\vec{v}$
are real vectors related to the  staggered magnetization 
$\langle\vec{S}_1-\vec{S}_2\rangle=2\vec{u}(1-\vec{u}^{2}-\vec{v}^{2})^{1/2}$
and \emph{vector chirality}
$\langle\vec{S}_1\times\vec{S}_2\rangle=\vec{v}(1-\vec{u}^{2}-\vec{v}^{2})^{1/2}$
of the dimer. Using the ansatz
$\vec{u}(\vec{r})=(-)^{\vec{r}}\vec{\varphi}(\vec{r})$, 
$\vec{v}(\vec{r})=(-)^{\vec{r}}\vec{\chi}(\vec{r})$,  
passing to the continuum in the coherent states path integral, 
and retaining up to quartic terms  in
$\vec{u}$, $\vec{v}$,
one
obtains the Euclidean action 
\begin{eqnarray} 
\label{Aeff} 
\mathcal{A}&=&\int d\tau d^{2}r \Big\{
\hbar(\vec{\varphi}\cdot \partial_{\tau}\vec{\chi}- \vec{\chi}\cdot\partial_{\tau}\vec{\varphi})\\
&+&(\vec{\varphi}^{2}+\vec{\chi}^{2})(J_{R}-3ZJ_{4}/2)
-Z[J_{\parallel}\vec{\varphi}^{2}+J_{4}\vec{\chi}^{2}] \nonumber\\
&+&(Z/2)[J_{\parallel}(\partial_{k}\vec{\varphi})^{2} +J_{4}(\partial_{k}\vec{\chi})^{2}]
+ZU_{4}(\vec{\varphi},\vec{\chi})\Big\}\nonumber,
\end{eqnarray}
where the quartic potential $U_{4}$ has the form
\begin{eqnarray} 
\label{U4} 
U_{4}&=&(\vec{\varphi}^{2}+\vec{\chi}^{2})[J_{\parallel}\vec{\varphi}^{2}
+J_{4}\vec{\chi}^{2}]/2\nonumber\\
&+&J_{4}(\vec{\varphi}^{2}+\vec{\chi}^{2})^{2} + (J_{\parallel}+J_{4})(\vec{\varphi}\times\vec{\chi})^{2}.
\end{eqnarray}
Interdimer interactions $J_{\parallel}$ and $J_{4}$ favor two competing
types of order: while $J_{\parallel}$ tends to establish the   
AF order ($\vec{\varphi}\not=0$, $\vec{\chi}=0$),
strong ring exchange $J_{4}$  favors another solution with 
 $\vec{\varphi}=0$, $\vec{\chi}\not=0$,
describing the state with a
\emph{staggered vector chirality}. It wins over the AF one for
$J_{4}> J_{\parallel}$,
$J_{4}>\frac{2}{5Z}J_{R}$, which for the square lattice ($Z=4$) translates into
\begin{equation} 
\label{svc} 
J_{4}>\max(J_{\parallel},J_{\perp}/9).
\end{equation}
On the line $J_{4}=J_{\parallel}$ 
the symmetry is enhanced from $\rm SU(2)$ to $\rm
SU(2)\times U(1)$, and the AF and chiral orders can coexist: a rotation $(\vec{\varphi}+i\vec{\chi})\mapsto
(\vec{\varphi}+i\vec{\chi})e^{i\alpha}$ leaves the 
action invariant.

The chiral state  may be viewed as an analog of the circulating
current state considered above: in terms of the original fermions of
the Hubbard model, the $z$-component of the chirality
$(\vec{S}_{1}\times\vec{S}_{2})_{z}= \frac{i}{2}\big\{ 
(a_{1\downarrow}^{\dag}a_{2\downarrow}^{\vphantom{\dag}})
(a_{2\uparrow}^{\dag}a_{1\uparrow}^{\vphantom{\dag}})
-(a_{2\downarrow}^{\dag}a_{1\downarrow}^{\vphantom{\dag}}) 
(a_{1\uparrow}^{\dag}a_{2\uparrow}^{\vphantom{\dag}})
\big\}$ corresponds to the \emph{spin current} (particles with up and down spins
moving in opposite directions).

\paragraph{Summary.--} 

I have considered fermionic and bosonic models on a bilayer
optical lattice which exhibit a phase transition into a
circulating current state with spontaneously broken
time reversal symmetry. The simplest of those models includes just
nearest-neighbor interactions and hoppings, and can possibly be
realized with the help of polar molecules.

\paragraph{Acknowledgments.--} I sincerely thank
 U.~Schollw{\"o}ck, T.~Vekua, and S.~Wessel for fruitful
discussions. Support by 
 Deutsche For\-schungs\-gemeinschaft (the Heisenberg Program, KO~2335/1-1) is
 gratefully acknowledged.

\end{document}